\documentclass[twocolumn,preprintnumbers,amsmath,amssymb]{revtex4}

\usepackage{graphicx}
\usepackage{dcolumn}
\usepackage{bm}
\usepackage{setspace}
\usepackage{amsfonts}
\usepackage{braket}
\newcommand{\Real}{\operatorname{Re}}
\newcommand{\Imag}{\operatorname{Im}}
\newcommand{\ff}{\mbox{\boldmath $f$}}

\newcommand{\uu}{\mbox{\boldmath $u$}}
\newcommand{\vv}{\mbox{\boldmath $v$}}

\newcommand{\ppsi}{\mbox{\boldmath $\psi$}}

\newcommand{\II}{\mbox{$\hat{I}$}}
\newcommand{\HH}{\mbox{$\hat{H}$}}
\newcommand{\MM}{\mbox{$\hat{M}$}}

\newcommand{\PP}{\mbox{$\hat{P}$}}

\begin{document}
\title{
Enhanced response of non-Hermitian photonic systems near exceptional points
 }
\author{
Satoshi Sunada
}
\affiliation{
Faculty of Mechanical Engineering, Institute of Science and Engineering,
Kanazawa University, Kakuma-machi Kanazawa, Ishikawa 920-1192, Japan 
}

\date{\today}
\begin{abstract}
This paper theoretically and numerically studies the response 
 characteristics of non-Hermitian resonant photonic systems operating near an 
exceptional point (EP), where two resonant eigenmodes coalesce. 
It is shown that a system near an EP can exhibit a non-Lorentzian
 frequency response, whose line shape and intensity strongly 
depend on the modal decay rate and coupling parameters for the input waves, 
unlike a normal Lorentzian response around a single resonance.     
In particular, it is shown that 
the peak intensity of the frequency response 
is inversely proportional to the fourth power of the modal decay rate and 
can be significantly enhanced with the aid of optical gain.
The theoretical results are numerically verified by a full wave simulation of a microring
 cavity with gain. 
In addition, the effects of the nonlinear gain saturation and spontaneous
 emission are discussed.
The response enhancement and its parametric dependence may be  
useful for designing and controlling the excitation of eigenmodes by external fields.  
\end{abstract}

\maketitle

\section{Introduction}
In a quantum system interacting with the surrounding environment, 
 the quantum property is described by a non-Hermitian
Hamiltonian formalism \cite{Okolowicz2003}.
The eigenvalues of the non-Hermitian Hamiltonian are generally complex, 
and the eigenmodes do not form
an orthogonal basis, unlike those of isolated (Hermitian) systems.  
Such non-Hermitian properties also appear in classical wave
systems such as acoustic or electromagnetic waves in open cavities. 

A remarkable feature of non-Hermitian systems is 
the nonorthogonality of the eigenmodes. 
When a non-Hermitian system is driven by noise, 
the nonorthogonality 
can lead to excess system noise whose amplitude 
can greatly exceed the level expected in Hermitian systems.  
To date, many theoretical and experimental works in different
contexts, including Bose--Einstein condensates, lasers, fluid dynamics, and
pattern formation, 
have revealed
that excess noise is a common feature of nonnormal (and non-Hermitian)
physical systems driven by noise
and manifests itself in certain forms \cite{
Trefethen2005,Trefethen1993,
Biancalani2017,Farrell1994,Szirmai2009,Papoff2008,Longhi2000,Alessandro2009}. 

In the context of laser physics, 
 excess spontaneous emission
noise in laser cavities has been well-studied 
\cite{Petermann1979,Haus1985,Siegman1989-1,Siegman1995,Exter2001,Cheng2006,SYLee2008}.
The excess noise has been observed as 
the broadening of the laser
linewidth \cite{Cheng1996,Eijkelenborg1996,Yao1992,vanderLee1997} 
and a low-frequency intensity fluctuation \cite{vanderLee2000}, 
and it has been characterized 
by the Petermann factor (PF), a measure of the nonorthogonality of
the eigenmodes. 
Importantly, the PF does not characterize the enhancement in the spontaneous
emission itself but rather 
the enhanced coupling of the spontaneous
emission to an eigenmode 
\cite{Exter2001,Cheng2006}.
This implies that the PF can be generalized as a factor characterizing
the enhanced response of an eigenmode to inputs, and 
the excess spontaneous emission is one aspect of the enhanced response. 
Actually, a system with a large PF can exhibit an excess excitation
response for external injection in an amplifier configuration 
\cite{Haus1985,Siegman1989-1,Siegman1995}.

It is known that the modal nonorthogonality is maximized and the PF diverges at a degeneracy
point called an exceptional point (EP), 
where both the eigenvalues and corresponding eigenmodes
coalesce \cite{Kato1966,Berry2004,Heiss2012,SYLee2008}.
Recently, EPs have become experimentally accessible in a variety of 
photonic systems \cite{SBLee2009,Cao2015,Zhen2015,Kim2016}
and have attracted much attention. 
In addition to the PF divergence, a number of unique properties related to EPs have been found
and demonstrated in the past few
years, such as  
asymmetric mode switching \cite{Doppler2016}, the reversal of the pump
dependence of a laser \cite{Liertzer2012} and the effect of loss
\cite{BoPeng2015}, 
nonreciprocal transmission \cite{BoPeng2014,LChang2014}, 
and unidirectional invisibility \cite{Lin2013,Feng2013}.  
Moreover, EPs have been used to enhance the sensitivity of microcavity sensors 
\cite{Wiersig2014,WChen2017,Hodaei2017,Ren2017,Sunada2017}.

In this paper, the response characteristics of non-Hermitian resonant
systems (optical cavities) near an EP are theoretically
and numerically studied.  
A linear response analysis near an EP
reveals that regardless of the PF,
the actual response intensity is limited to a finite value.
Instead, a variety of cavity responses to inputs are exhibited near an EP, 
mainly depending on the modal decay rates and coupling parameters for the input waves.
In particular, when an optical cavity with gain operates at an EP, 
the response intensity can be excessively enhanced with the aid of the gain. 
The condition for the enhancement is derived and discussed. 
These theoretical results are numerically verified by a dynamical
model describing the interaction between the light field and a two-level 
gain medium.  
It is also discussed that the nonlinear gain saturation and spontaneous emission
limit the enhancement in the response intensity and quality. 

The rest of this paper is organized as follows. In Sec. \ref{secII}, 
the effect of the nonorthogonality of the eigenmodes on 
the system response is briefly introduced.
Then, a general expression of the system response at an EP is 
provided in a simple 2$\times$2 matrix form, and 
the frequency responses are analyzed. 
In Sec. \ref{secIII}, 
the theoretical results 
are numerically
verified in a full wave simulation of a microring cavity operating near
an EP.
Finally, a summary is provided in Sec. \ref{secIV}. 

\section{Linear Responses of Non-Hermitian Systems near an EP \label{secII}}
\subsection{Model and Petermann factor \label{secIIa}}
We consider an optical cavity system driven by an input field 
and analyze the response characteristics 
in the form of a coupled mode theory.
First, suppose that an optical cavity possesses $n$ modes coupled with
each other and that it is described by an $n\times n$ non-Hermitian (effective Hamiltonian)
matrix $\HH$,
which represents the resonances and modal coupling. 
The non-Hermiticity of $\HH$ arises from the radiation loss, absorption loss,
and gain inside the cavity. 
Then, we also suppose that the slowly varying envelope of the intracavity
optical field is characterized by an $n$-dimensional state vector $\ppsi
\in \mathbb{C}^n$, and it is excited by an input field, denoted by 
$\ff \in \mathbb{C}^n$. 
The time evolution of $\ppsi$ driven by $\ff$ is generally given by 
\begin{eqnarray}
i\dfrac{d\ppsi}{dt} = \HH\ppsi + \ff. \label{eq1}
\end{eqnarray}
This coupled mode equation can be obtained from the Maxwell equations by 
assuming that the optical field varies slowly in time with respect to a
reference frequency and it is expanded by appropriate basis functions. 
In this study, we are interested in how the state vector $\ppsi$
responds to an input $\ff$
because typical cavity properties such as reflection or transmission
can be characterized by the response $\ppsi$. 

A common way to analyze Eq. (\ref{eq1}) is an
expansion by the eigenmodes of $\HH$.
Although the eigenmodes of a non-Hermitian matrix are generally not 
orthogonal, 
this drawback is covered by the biorthogonality between the left and
right eigenmodes. 
Here, suppose that $\uu_j$ and $\vv_j$ are the right and
left eigenmode vectors of $\HH$,
respectively, where $j$ is a mode number $(j\in \{1,\cdots,n\})$. 
$\uu_j$ and $\vv_j$ are defined as 
$\HH\uu_j=\Omega_j\uu_j$ and
$\vv_j^{\dagger}\HH=\Omega_j\vv_j^{\dagger}$ with the eigenvalue
$\Omega_j \in \mathbb{C}$, where $\dagger$ denotes the Hermitian conjugate, 
and $\vv_i^{\dagger}\cdot\uu_j = 0$ 
$(i\ne j)$. 
The eigenvalue $\Omega_j = \omega_j - i\gamma_j$ represents a complex-valued
eigenfrequency of mode $j$ in the cavity; 
$\omega_j \in \mathbb{R}$ represents the resonant frequency, 
whereas $\gamma_j \in \mathbb{R}$ represents the decay (growth) rate
if it has a positive (negative) value. 
In this paper, we consider only $\gamma_j >0$ for
all $j$, i.e., all decaying modes. 

By expanding $\ppsi$ by the right eigenmodes as 
$\ppsi = \sum_j a_j(t)\uu_j$ and using the biorthogonal relation in Eq. (\ref{eq1}),
we obtain the mode equations, 
\begin{eqnarray}
i\dfrac{d a_j}{dt} = \Omega_ja_j + f_j, \label{eq2}
\end{eqnarray}
and the solution after a long time, 
$
a_j(t) = -i\int^t e^{-i\Omega_j(t-\tau)}f_j(\tau)d\tau,
$
where $f_j = \vv_j^{\dagger}\cdot \ff/(\vv_j^{\dagger}\cdot\uu_j)$.
Importantly, the amplitude $a_j$ is determined by $f_j$, and the
magnitude $|f_j|$ can be expressed as 
$
\sqrt{K_j}
|
\vv_j^{\dagger}\cdot\ff
|
/
(\|\uu_j\|\|\vv_j\|)
$, 
where 
$K_j = (\|\uu_j\|^2\|\vv_j\|^2)/
|
\vv_j^{\dagger}\cdot\uu_j 
|^2
$, and 
$\|\cdot\|$ denotes the usual Euclidean vector norm.
$K_j$ 
characterizes the coupling to mode $j$, 
and it is associated with
the condition number of the eigenvalue $\Omega_j$, i.e., the sensitivity to
perturbations \cite{Trefethen2005,Demmel1997} 
and the PF %
in the context of laser physics \cite{Berry2004,vanderLee1997}. 
$K_j \ge 1$ is always satisfied according to the
Cauchy--Bunyakovsky--Schwarz inequality. 
In particular, $K_j >1$ when $\HH$ is nonnormal
($\HH\HH^{\dagger}\ne\HH^{\dagger}\HH$) \cite{Comment11}.
Interestingly, $K_j$ diverges 
just at an EP, where 
$\uu_j$ completely overlaps another eigenvector, e.g.,
$\uu_{j'}$ ($j'\ne j$), 
because of the self-orthogonality  
$
\vv_j^{\dagger}\cdot\uu_j = \vv_{j}^{\dagger}\cdot\uu_{j'} = 0
$ \cite{SYLee2008}. 
However, $K_j$ is no longer valid at the EP because 
the eigenmode expansion breaks at the point. 
In other words, any $n$-dimensional state vector cannot be represented by 
the coalescing eigenmode basis at the EP.
The basis of the expansion can be completed by introducing additional 
vectors, i.e., the associated vectors defined by the Jordan chain
relations \cite{Seyranian2003,Seyranian2005}. 

\subsection{Frequency response at an EP \label{secIIb}}
As a starting point for deriving the cavity responses at an EP, 
we consider them in the frequency domain.  
By Fourier-transforming Eq. (\ref{eq1}) with respect to the time $t$,
we obtain
\begin{eqnarray}
\tilde{\ppsi}(\omega)
=
\MM(\omega)\tilde{\ff}(\omega),  
\label{eq_resolvent}
\end{eqnarray}
where $\tilde{\ppsi}(\omega)$ and $\tilde{\ff}(\omega)$ denote the Fourier
transforms of $\ppsi(t)$ 
and $\ff(t)$, respectively. 
$\MM(\omega) = (\omega\II-\HH)^{-1}$ is the resolvent of
$\HH$ and characterizes the response to an input wave 
with a frequency of $\omega$. $\II$ is an identity matrix. 

Then, we consider an optical cavity operating 
near a (second-order) EP, where only two eigenmodes of $\HH$ coalesce. 
When the frequency $\omega$ is close to the resonant frequencies of the two
eigenmodes and the influence of the other $n-2$ modes is sufficiently
weak, 
the $n-$dimensional matrix problem can be essentially reduced to a
two-dimensional matrix problem near the EP. 
By describing an effective 2$\times$2 Hamiltonian matrix at an EP as
$\HH_0$ and using
an expansion method based on the Jordan chain relation \cite{Pick2017}, 
the resolvent $\MM_{EP}$ at the EP is given by 
\begin{eqnarray}
\MM_{EP}(\omega)
=
R_0(\omega)\II + R_0^2(\omega)(
\HH_0-\Omega_0\II) \nonumber \\
=
\left(
\begin{array}{rr}
R_0(\omega) + c_{11}R_0^2(\omega), & c_{12}R_0^2(\omega)\\
c_{21}R_0^2(\omega), & R_0(\omega) + c_{22}R_0^2(\omega)
\end{array}
\right), \label{eq_EPres}
\end{eqnarray}
where
$R_0(\omega) = (\omega-\Omega_0)^{-1}$, and $\Omega_0$ is an 
eigenvalue of $\HH_0$ at the EP.
Further, $c_{ij}$ is the $ij$ component of the matrix
$(\HH_0-\Omega_0\II)$.
The derivation of Eq. (\ref{eq_EPres}) is shown in Appendix \ref{appen1}. 

A remarkable feature of the resolvent $\MM_{EP}$ is the presence of 
the second-order pole $R_0^2$, which appears only near the EP. 
From Eqs. (\ref{eq_resolvent}) and (\ref{eq_EPres}), the $i$ component of the field
vector, $\tilde{\psi}_i(\omega)$, driven by
$\tilde{\ff}(\omega)=(\tilde{f}_1,\tilde{f}_2)^T$, is 
$\tilde{f}_iR_0+\delta_i R_0^2$, 
where $\delta_i = \sum_jc_{ij}\tilde{f}_j$. 
Interesting behavior can be observed for $\gamma_0 \sim
|\delta_i/\tilde{f}_i|$ or $\gamma_0 \ll   |\delta_i/\tilde{f}_i|$, where 
$\gamma_0 =-\Imag\Omega_0$. 
 In the former case, the interference of the two terms
can lead to asymmetric response behavior with respect to
the resonant frequency, $\omega_0=\Real\Omega_0$,  
or suppression of the response amplitude,
such as Fano--Feshbach resonances \cite{WDHeiss2014,Suh2004}.
On the other hand, in the latter case ($\gamma_0 \ll   |\delta_i/\tilde{f}_i|$), 
$\tilde{\psi}_i$ can be approximated as $\delta_i R_0^2$ near the resonant frequency 
$\omega_0$;
thus, the intensity $|\tilde{\psi}_i|^2$ has a squared Lorentzian
shape, i.e., $|\delta_iR_0^2|^2 =
|\delta_i|^2/[\gamma_0^2+(\omega-\omega_0)^2]^2$, with a peak
intensity proportional to $\gamma_0^{-4}$, 
whereas the peak intensity of 
a standard Lorentzian shape is proportional to $\gamma_0^{-2}$.  
Although $\gamma_0$ is generally dependent on $c_i$ and the input wave
couplings in passive cavities, it can be changed by optical gain in
active cavities.  
Therefore, when $\gamma_0$ is reduced by the gain, the response intensity near the resonance
can be greater than the standard Lorentzian-type
responses in cavities that do not
operate at an EP.  

It has been reported that a similar enhancement appears for 
spontaneous emission \cite{Lin2016,Yoo2011,Pick2017}.
The above discussion suggests that the enhanced spontaneous emission is one aspect of
the enhanced cavity responses; a cavity operating at an
EP can respond strongly to other inputs. 

\subsection{Example of enhanced response \label{secIIc}}
\begin{figure}[t]
\begin{center}
  \begin{tabular}{c}
\hspace*{-1cm}
\raisebox{0.0cm}{\includegraphics[width=6cm]{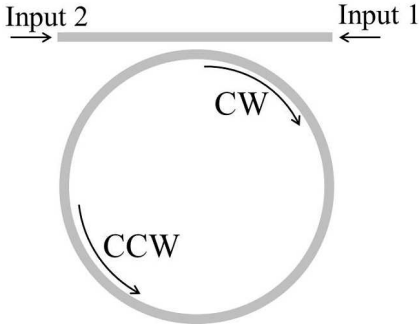}}
  \end{tabular}
\end{center}
\vspace{-4mm}
\caption{\label{figure1} 
Microring cavity coupled to a straight waveguide.  
The input waves from the left and right ports in the straight waveguide 
couple to the CW and CCW modes in the microring, respectively. 
The two modes are also coupled to each other by backscattering. 
} 
\end{figure}

As an example of an optical cavity that can operate at an EP, 
we choose a microring cavity with non-Hermitian
backscattering \cite{Peng2016,Wiersig2011,Shu2016} 
and analyze the cavity response to an incident wave with
a frequency of $\omega$. 
The wave is coupled to the cavity via a waveguide [see
Fig. \ref{figure1}]. 
In the cavity, the wave can propagate in the clockwise (CW) or
counterclockwise (CCW) direction along the ring waveguide.  
The time evolution of the intracavity field in the CW and CCW 
traveling wave basis is given by 
 \begin{eqnarray}
i\dfrac{d}{dt}
\left(
\begin{array}{c}
a_{1} \\
a_{2}
\end{array}
\right)
=
\left(
\begin{array}{cc}
\Omega_0 & -\epsilon /2 \\
-p/2, & \Omega_0 
\end{array}
\right)
\left(
\begin{array}{c}
a_{1} \\
a_{2}
\end{array}
\right)
+
\left(
\begin{array}{c}
\kappa_{1}  \\
\kappa_{2} 
\end{array}
\right) e^{-i\omega t}, \label{eq_ring}
\end{eqnarray}
where $a_{1}$ and $a_{2}$ are the amplitudes of the CCW and CW waves, respectively.
$\kappa_1$ and $\kappa_2$ represent the coupling of the incident wave to
the CCW and CW waves, respectively.  
$\Omega_0 = \omega_0-i\gamma_0$ is the eigenfrequency of the ring cavity
modes 
(CCW and CW modes) when there are no coupling terms,
i.e., $\epsilon = p = 0$. 
$\epsilon$ ($p$) represents the backscattering coupling from the 
CW (CCW) wave to the CCW (CW) wave. 
In general, the magnitudes and phases of $\epsilon$ and $p$ can be 
controlled by placing nanoscatterers near the cavity 
\cite{Peng2016}, making the cavity geometry asymmetric \cite{Wiersig2011}, 
or introducing modulations in the refractive index and
dissipation inside a cavity \cite{Feng2013,Shu2016}.

Backscattering coupling typically causes the splitting of the degenerate 
eigenvalues of the CW and CCW modes. 
However, when the backscattering couplings are highly asymmetric, i.e., 
$\epsilon \ne 0$ and $p$ = 0 or $\epsilon =0$ and $p\ne 0$, 
the split eigenvalues coalesce into a
single value \cite{Wiersig2011}. 
(In this cavity, the EPs are distributed along the line $p=0$ or
$\epsilon=0$ except for the origin, $\epsilon=p=0$, in the $\epsilon$--$p$
parametric plane.)
The coalesced eigenvalue is $\Omega_0$ at an EP, and the corresponding
eigenmode is the CCW (CW) mode when $\epsilon\ne 0$ and $p=0$
($\epsilon =0$ and $p\ne 0$).  
This non-Hermitian degeneracy is different from the degeneracy in
ring cavities without backscattering, i.e., $\epsilon = p =0$, 
where there are two linearly independent eigenmodes (the CW and CCW modes)
with the same eigenfrequency $\Omega_0$. 
The normal degeneracy point has been referred to as 
the diabolic point (DP). 

At an EP ($\epsilon \ne 0$ and $p =0$),
the CW and CCW wave amplitudes in the frequency domain are represented by 
\begin{eqnarray}
\left(
\begin{array}{c}
\tilde{a}_{1}(\omega) \\
\tilde{a}_{2}(\omega)
\end{array}
\right)
=
\dfrac{1}{\omega - \Omega_0}
\left(
\begin{array}{cc}
1, & 
-\dfrac{\epsilon}{
2\left(
\omega - \Omega_0
\right)
}\\
0,
&
1
\end{array}
\right)
\left(
\begin{array}{c}
\kappa_1 \\
\kappa_2
\end{array}
\right). \label{eq_microring}
\end{eqnarray}
The CCW wave amplitude, $\tilde{a}_1$, is affected by the backscattering $\epsilon$, 
decay rate $\gamma_0$, and coupling terms $\kappa_1$ and
$\kappa_2$.   
When a wave with a frequency of
$\omega \approx \omega_0$ is incident upon the cavity in the CW direction,
i.e., $\kappa_{1} = 0$ and $\kappa_{2} \ne 0$, 
the two-mode intensities near the resonance
$\omega \approx \omega_0$ are given by 
$|\tilde{a}_{1}(\omega_0)|^2 \approx |\epsilon \kappa_2|^2/(4\gamma_0^{4})$ 
and $|\tilde{a}_{2}(\omega_0)|^2 \approx |\kappa_2|^2 \gamma_0^{-2}$. 
Thus, 
$|\tilde{a}_{1}(\omega_0)|^2 \gg |\tilde{a}_{2}(\omega_0)|^2$
for $\gamma_0 \ll |\epsilon |/2 (=\gamma_s)$; that is, the CCW wave is strongly excited by 
the incident wave in the CW direction. 
This counterintuitive response does not arise 
in a ring cavity without backscattering ($\epsilon=p=0$) 
or in cavities with isolated resonances.
Numerical verification is presented in Sec. \ref{secIII}. 

\subsection{Transient growth and excess noise \label{secIId}}
\begin{figure}[t]
\begin{center}
  \begin{tabular}{c}
\hspace*{-0.4cm}
\raisebox{0.0cm}{\includegraphics[width=9cm]{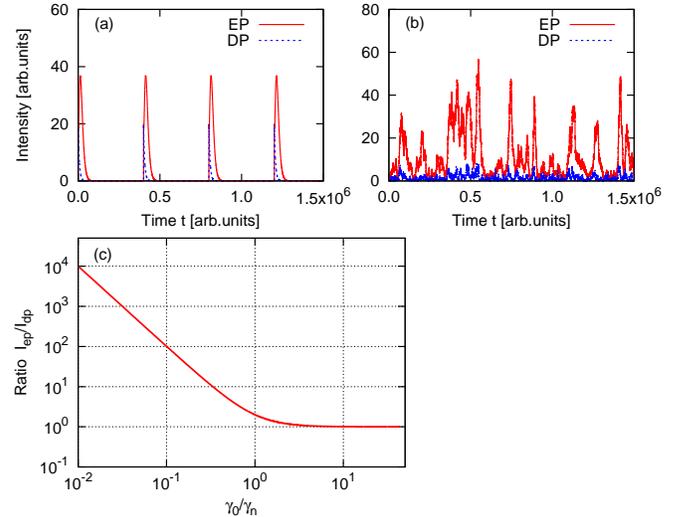}}
  \end{tabular}
\end{center}
\vspace{-4mm}
\caption{\label{figure4}
(a), (b) Dynamics of the intensity $\|\ppsi\|^2 =\sum_j|a_j(t)|^2$ at
 the EP ($\epsilon \ne 0$ and $p=0$) in a microring cavity under
 (a) a delta-function-like pulse with a period 
$T=4\times 10^5$ 
 and (b) a Gaussian noise input (red curves).
These results were obtained from Eq. (\ref{eq1}) with
the $2\times 2$ matrix of the microring cavity for $\gamma_0/\gamma_n=0.4$. 
For comparison, the intensity in a ring cavity without
 backscattering, 
i.e., a cavity at the DP ($\epsilon=p=0$), 
is also shown as the blue dotted curve for each case. 
In (b), $\ff=(f_1,f_2)^T$ is treated as complex white Gaussian noise,
$\braket{f_i^*(t')f_j(t)}=2D\delta_{ij}\delta(t-t')$, where $D$ is the
 noise variance. 
(c) $\gamma_0/\gamma_n$ dependence of $I_{ep}/I_{dp}$. 
$I_{ep}$ indicates the time average of the response intensity dynamics at the
 EP,
whereas $I_{dp}$ indicates that at the DP ($\epsilon=p=0$). 
Gaussian noise is applied to the cavity. 
}
\end{figure}

Because the cavity responses are characterized by Eqs.
(\ref{eq_resolvent}) and (\ref{eq_EPres}), 
cavities operating near EPs can respond sensitively to not only
monochromatic inputs but also other inputs
[see Figs. \ref{figure4}(a) and (b) for examples of the responses to pulsed and
random inputs].
From Eq. (\ref{eq_resolvent}), we note that 
in the presence of internal noise such as spontaneous emission,
the cavities can also be sensitive to the noise. 
This is excess noise in the cavity and limits the response quality. 

When we consider the response properties in the time domain,   
we note that the impulse response corresponding to $R_0^2$ in
Eq. (\ref{eq_EPres}) 
is given by $t\exp(-i\Omega_0 t)$, whose amplitude grows transiently 
on a short time scale before decaying.
As seen in Figs. \ref{figure4}(a) and (b), 
the transient growth is sustained by sequential pulses or noise stimuli.
Excess noise can be characterized as the lasting transient growth due to
internal noise. 
According to nonnormal operator theory \cite{Farrell1996}, transient
growth can dominate the dynamics of the intensity $\|\ppsi\|^2$ if 
at least one eigenvalue of $i(\HH^{\dagger}-\HH)$ (the non-Hermitian
part of $\HH$) is positive. 
In the 2$\times 2$ matrix model at an EP, the condition is given by 
\begin{eqnarray}
\gamma_0 < \gamma_n = \dfrac{1}{2}\sqrt{
-\left(
c_{11} + c_{22}^*
\right)^2
+ \left|
c_{21}^*-c_{12}
\right|^2
}. \label{eq_con}
\end{eqnarray}

In a microring cavity at the EP ($\epsilon \ne 0$ and $p=0$), 
the condition is $\gamma_0 < \gamma_n = |\epsilon |/4$.
Figure \ref{figure4}(c) shows the appearance of excess noise 
in a microring cavity, where
 the time-averaged response intensity 
to Gaussian noise at an EP ($\epsilon\ne 0$ and $p=0$), represented by 
$I_{ep}$,
is compared to  
that of the response intensity at a DP ($\epsilon=p=0$), 
$I_{dp}$.
When $\gamma_0 \ll \gamma_n$,  
$I_{ep}$ is $\gamma_n^2/\gamma_0^2$ times greater than 
$I_{dp}$.

\section{Full Wave Simulations \label{secIII}}
\subsection{Dynamical model}
In this section, we numerically check the validity of the 
response enhancement presented in the previous section.
This numerical verification is important because 
the linear response model is based on a simple (2$\times$2 matrix)
model, i.e., a two-mode approximation near an EP; 
however, in realistic systems, multiple modes may be involved. 
Moreover, to reduce the decay rate $\gamma_0$ and achieve an
excess enhancement in the response intensity, loss compensation by
optical gain is needed. 
However, amplification by the gain is inevitably accompanied by spontaneous
emission noise. 
In addition, an actual gain material is nonlinear and saturates the amplified
light intensity. 
Therefore, the numerical simulations are conducted using a
dynamical model incorporating these effects. 
In this paper, we use a model describing the dynamics of
the slowly varying envelope $E$ of the electric field inside a cavity, the 
polarization field $\rho$, and the population inversion component $W$ in
a two-level gain medium \cite{Harayama2005,Sunada2013}:
\begin{eqnarray}
\dfrac{\partial E}{\partial t}
= \dfrac{i}{2}\left[
\dfrac{\partial^2}{\partial x^2}
+ 
\dfrac{n^2(x)}{n_0^2}
\right]E
+
\xi \rho + E_{in},  \label{SB1} \\ 
\dfrac{\partial \rho}{\partial t}
=
-
\left(
\gamma_{\perp} 
+
i\Delta_a
\right)
\rho
+
\gamma_{\perp} 
WE
 + F_{1}, \label{SB2} \\
%
\dfrac{\partial W}{\partial t}
=
-
\gamma_{\parallel} 
\left(W-W_{\infty}
\right)
- 2\gamma_{\parallel}
\left(
E\rho^* + E^*\rho
\right)+ F_{2}, \label{SB3}
\end{eqnarray}
where space and time are made dimensionless by the scale transformations
$n_0\omega_sx/c \rightarrow x$ and $\omega_st \rightarrow t$,
respectively. 
$\omega_s$ is a reference frequency close to the transition frequency
$\omega_a$ of
the two-level gain medium. 
In Eqs. (\ref{SB1})--(\ref{SB3}), $E$, $\rho$, $W$, and all of the other parameters
are also made dimensionless. 
Further, $n$ is the refractive index inside the cavity, 
$n_0$ is the spatially averaged refractive index,  
$\xi = 2\pi/n_0^2$ is a coupling constant, 
and
$\Delta_a$ represents the gain center. 
(The relationship between $\Delta_a$ and the actual transition frequency $\omega_a$ is
given by $\Delta_a = \omega_a/\omega_s-1$.)
The two relaxation parameters, $\gamma_{\perp}$ and
$\gamma_{\parallel}$, are the
transverse and longitudinal relaxation rates,
respectively. 
$W_{\infty}$ represents the pumping power, which is effectively used to 
reduce the cavity loss.  

In Eq. (\ref{SB1}), $E_{in}$ represents the input field, which is coupled
to the intracavity field $E$.
$F_1$ and $F_2$ represent spontaneous emission noise from the
gain medium, 
and they are modeled as complex white Gaussian noise \cite{Drummond1991,Cerjan2015}.  
The specific forms of $F_1$ and $F_2$ are similar to those 
reported in \cite{Cerjan2015}. 

\subsection{Cavity model and parameters}
For verification, we choose a microring cavity, which is 
discussed in Sec. \ref{secIIc}. 
The cavity is modeled as a one-dimensional 
ring waveguide with a length of $L$. 
A periodic boundary condition 
is imposed on the intracavity field as $E(x,t) = E(x+L,t)$,   
where $x$ denotes the coordinate along the ring waveguide. 
Tuning to an EP in the ring cavity is possible by modulating the
complex-valued refractive index $n^2(x)$ inside the cavity \cite{Shu2016}:
\begin{eqnarray}
n^2(x) = n_0^2\left[1 + \epsilon \exp({2ik_0x})  + p \exp({-2ik_0x}) + 2i\beta \right], 
\end{eqnarray}
where $k_0 = 2\pi m_0/L$ ($m_0$ is an integer) represents the 
resonant wavenumber of the cavity for $\epsilon=p=0$, 
and $\beta$ represents the absorption loss rate.  
The refractive index modulation induces strong linear coupling between
the nearly degenerate eigenmodes with the $+k_0$ (CCW)
and $-k_0$ (CW) wave components. 
In particular, when $\epsilon \ne 0$ and $p=0$, the two eigenmodes
collapse to a single mode, which can be expressed as 
the CCW mode of the wavenumber $k_0$.  
Because the physical meanings of $\epsilon$ and $p$ are the same as 
those described in Sec. \ref{secIIc}, we use the same notation.  

In this simulation, the following parameters are fixed: 
$L/(2\pi) =10$, $n_0=3.0$, $k_0=1 (m_0=10)$, $\xi=2\pi/n_0^2$,
$\beta/\xi=10^{-3}$, $\Delta_a=0$, $\gamma_{\perp}=0.1$, and $\gamma_{\parallel}=10^{-3}$. 
For these parameters, the resonant frequency $\Delta_0$
corresponding to the wavenumber $k_0$ at the EP is set to be equal 
to the gain center $\Delta_a$, 
and the modes with $\Delta_0$ are selectively pumped. 
The effective decay rate, including the effect of the gain, 
is defined as $\gamma_0 = \beta - \xi W_{\infty}$ \cite{Harayama2005}
and is changed by $W_{\infty}$.   
In this paper, $W_{\infty}$ is kept below the threshold pumping power $W_{th}=\beta/\xi$, 
and $\gamma_0$ is always positive. 

\begin{figure}[t]
\begin{center}
  \begin{tabular}{c}
\hspace*{-0.4cm}
\raisebox{0.0cm}{\includegraphics[width=8.8cm]{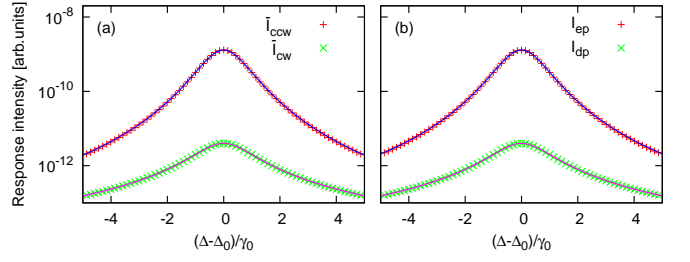}}
  \end{tabular}
\end{center}
\vspace{-4mm}
\caption{\label{fig_freq1} 
(a) 
Time-averaged response intensities of the CCW and CW waves 
at the EP ($\epsilon=1.8\beta$ and $p$=0),  
$\bar{I}_{ccw}$ and $\bar{I}_{cw}$, 
under the CW wave input condition ($\kappa_1=0$,
 $\kappa_2=10^{-7}\beta$). 
$\Delta_0$ is the degenerate resonant frequency at the EP.  
The theoretical curves of the CCW and CW wave intensities 
obtained from Eq. (\ref{eq_microring}) are shown as the blue and pink
 solid curves, respectively. 
(b) Intensity $I_{ep}=\bar{I}_{ccw}+\bar{I}_{cw}$ at the EP
under the same CW wave input condition. 
For comparison, the intensity $I_{dp}$ 
at the DP ($\epsilon=p=0$) is also shown. 
The theoretical curves obtained from Eq. (\ref{eq_microring}) 
at the EP and DP are shown as the blue and pink solid curves, respectively. 
In (a) and (b), $W_{\infty}=0.95\times
 10^{-3}$, and $\gamma_0/\gamma_s=1/18$. 
}
\end{figure}

\begin{figure}[t]
\begin{center}
  \begin{tabular}{c}
\hspace*{-0.4cm}
\raisebox{0.0cm}{\includegraphics[width=8.8cm]{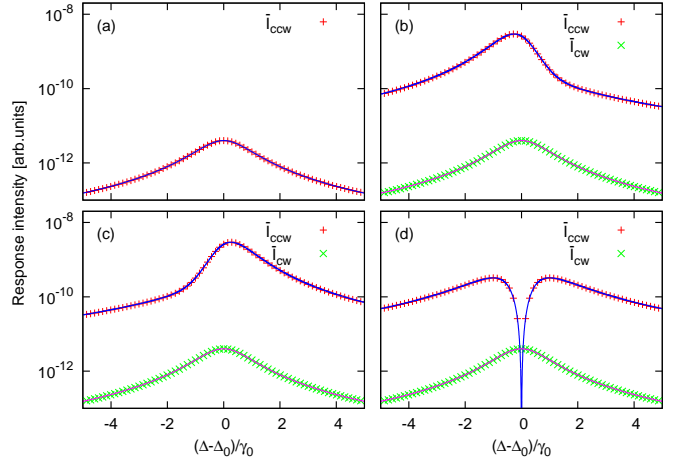}}
  \end{tabular}
\end{center}
\vspace{-4mm}
\caption{\label{fig_freq2} 
(a) Time-averaged response intensity $\bar{I}_{ccw}$ under the
 CCW wave input condition ($\kappa_1=10^{-7}\beta$ and $\kappa_2=0$).
(b)--(d) Time-averaged response intensities $\bar{I}_{ccw}$ and $\bar{I}_{cw}$ under
 a bidirectional wave input condition with 
(b)
$\kappa_1=\epsilon/(2\gamma_0)\kappa_2$, 
(c) $\kappa_1=-\epsilon/(2\gamma_0)\kappa_2$, 
and 
(d)
$\kappa_1=-i\epsilon/(2\gamma_0)\kappa_2$, where
 $\kappa_2=10^{-7}\beta$. 
In (a)--(d), $\epsilon =1.8\beta$, $p=0$, $W_{\infty}=0.95\times
 10^{-3}$, and $\gamma_0/\gamma_s=1/18$. 
The theoretical curves of the CCW and CW wave intensities obtained from Eq. (\ref{eq_microring}) for each
 input condition are shown as the blue and pink solid curves,
 respectively.
}
\end{figure}

\begin{figure}[t]
\begin{center}
  \begin{tabular}{c}
\hspace*{-0.4cm}
\raisebox{0.0cm}{\includegraphics[width=7cm]{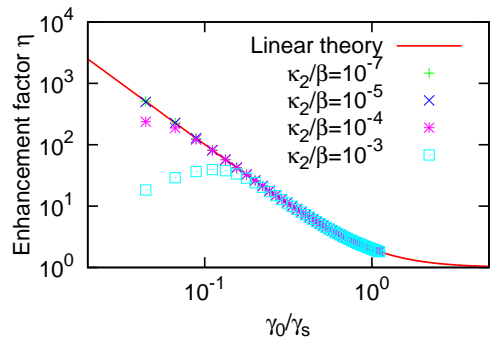}}
  \end{tabular}
\end{center}
\vspace{-4mm}
\caption{\label{cw-winf} 
Enhancement factor $\eta$ versus $\gamma_0/\gamma_s$. 
$\eta$ is measured under the CW wave input condition ($\kappa_1=0$) at the resonance
 $\Delta=\Delta_0$.
The theoretical curve is given by $\eta = \gamma^2_{s}/\gamma_0^2 +
 1$.   
}
\end{figure}

\begin{figure}[t]
\begin{center}
  \begin{tabular}{c}
\hspace*{-0.4cm}
\raisebox{0.0cm}{\includegraphics[width=8.8cm]{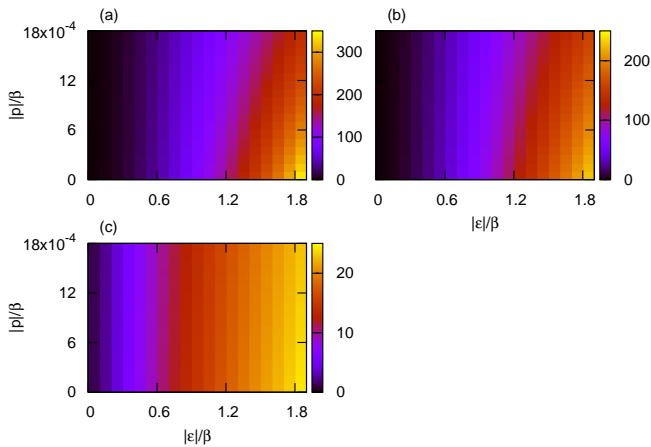}}
  \end{tabular}
\end{center}
\vspace{-4mm}
\caption{\label{epplot}
Enhancement factor $\eta$ plotted in the $\epsilon$--$p$ plane. 
(a) $\kappa_2=10^{-7}\beta$, (b) $\kappa_2=10^{-4}\beta$, and
(c) $\kappa_3=10^{-3}\beta$.
In (a)--(c), $W_{\infty}=0.95\times 10^{-3}$, and $\arg(\epsilon/p)=0$.}
\end{figure}

\begin{figure}[t]
\begin{center}
  \begin{tabular}{c}
\hspace*{-0.4cm}
\raisebox{0.0cm}{\includegraphics[width=6cm]{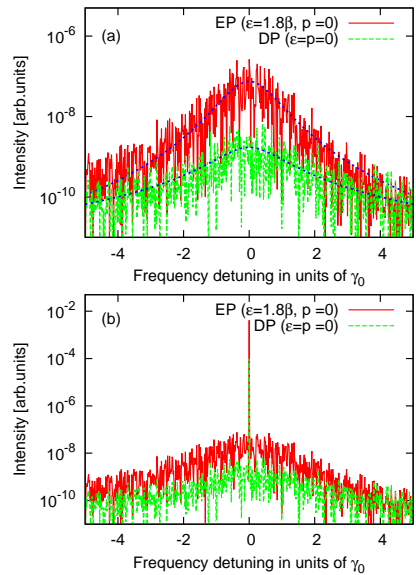}}
  \end{tabular}
\end{center}
\vspace{-4mm}
\caption{\label{fig6} 
(a) Each emission spectrum at the EP
 ($\epsilon=1.8\beta$ and $p=0$) and DP ($\epsilon=p=0$). 
The spectrum at the EP is fitted by a squared Lorentzian curve, whereas 
the spectrum at the DP is fitted by a Lorentzian curve. 
The fitting curves are shown as the blue dotted curves.  
(b) Emission spectra when an input wave with a frequency of
 $\Delta_0$ is coupled to the cavity in the CW direction. 
In (a) and (b), 
$\kappa$ = $10^{-3}\beta$, $W_{\infty}=0.9\times 10^{-3}$, and 
$\gamma_0/\gamma_n=2/9$. 
}
\end{figure}

\subsection{Response characteristics}
In this subsection, for simplicity, we omit the noise terms, $F_1$ and
$F_2$, 
and we consider the case in which two input waves with a frequency of $\Delta$ are coupled 
to the cavity, as shown in Fig. \ref{figure1}, i.e., 
$E_{in} = \kappa_1 \exp(ik_0x -i\Delta t) + \kappa_2 \exp(-ik_0x -i\Delta t)$, 
where
$\kappa_1$ and $\kappa_2$ represent the CCW and CW
wave components of the input waves in the cavity, respectively.  
To analyze the responses to the input waves,
we measure the intensities of the CCW and CW wave components of the
field $E$ as  
$I_{ccw} =  \sum_{m>0}|a_m|^2$ and $I_{cw}=\sum_{m<0}|a_m|^2$,
respectively, 
where 
$a_m = 1/L\int_0^L E(x,t) e^{-ik_m x} dx$, and $k_m=2\pi m/L$ ($m \in
\mathbb{Z}$), 
and calculate the time averages of $I_{cw}$ and $I_{ccw}$ after
relaxation to a steady state. 

Here, we consider the responses at the EP ($\epsilon\ne 0$ and $p=0$)
under the CW wave input condition, i.e., $\kappa_1=0$ and
$\kappa_2\ne 0$.
In Fig. \ref{fig_freq1}(a), 
the time-averaged intensities $\bar{I}_{ccw}$ and $\bar{I}_{cw}$ under the
input condition are plotted as a function of the frequency $\Delta$. 
$\epsilon$ and $W_{\infty}$ are set so that $\gamma_0$ is smaller than 
the critical value $\gamma_s
(=|\epsilon |/2)$, which is provided in Sec. \ref{secIIc}. 
For a relatively low input intensity, i.e., $\kappa_2=10^{-7}\beta$, 
the numerical results correspond well to the theoretical results
obtained from Eq. (\ref{eq_microring});
the frequency response intensities of the CCW and CW waves 
have a squared Lorentzian shape and conventional Lorentzian shape, respectively.
We can see that the CCW wave intensity 
at the resonance ($\Delta =\Delta_0$)
is two orders of magnitude greater than the CW wave intensity [see Fig. \ref{fig_freq1}(a)]. 

According to Eq. (\ref{eq_microring}), the response characteristics at
an EP can be controlled by varying the input waves. 
For example, under the CCW wave input condition ($\kappa_1\ne 0$ and
$\kappa_2=0$), the CCW wave 
intensity is not enhanced, as shown in Fig. \ref{fig_freq2}(a), 
and the frequency response intensity has a Lorentzian shape.  
On the other hand, when two input waves with well-controlled phases and
amplitudes are simultaneously coupled to the cavity, 
the interference of the two input waves leads to an asymmetric response
curve for $\kappa_1=\pm\epsilon/(2\gamma_0)\kappa_2$ 
[Figs. \ref{fig_freq2}(b) and \ref{fig_freq2}(c)] or suppression of the response
intensity at the resonance ($\Delta =\Delta_0$) for $\kappa_1=-i\epsilon/(2\gamma_0)\kappa_2$ 
[Fig. \ref{fig_freq2}(d)].
This suggests the possibility of novel mode switching; however, 
its further study is beyond the scope of this paper.
\subsection{Response enhancement and its limitations}
In what follows, we consider only the CW wave input condition 
($\kappa_1=0$ and $\kappa_2\ne 0$) and compare the
response intensity at the EP to that in a ring cavity with the same
radius, the same loss rate $\gamma_0$, and the same coupling strength
$\kappa_2$ but without non-Hermitian backscattering, 
i.e., the response intensity at the DP ($\epsilon=p=0$). 

Figure \ref{fig_freq1}(b) shows an example that compares
the frequency response intensity at the EP, $I_{ep}$, and that at the DP,
$I_{dp}$, which are calculated as $\bar{I}_{ccw}+\bar{I}_{cw}$. 
To quantitatively evaluate the enhancement, 
the enhancement factor $\eta$ was defined as the ratio of each peak 
intensity at the resonance ($\Delta=\Delta_0$), i.e., $\eta = I_{ep}/I_{dp}$.
For $\kappa_2=10^{-7}\beta$ and $\gamma_0/\gamma_s=1/18$, 
we obtain $\eta \approx 325$. 

The parametric dependence of $\eta$ is summarized in Figs. \ref{cw-winf} and \ref{epplot}. 
For a low input intensity, i.e., 
$\kappa_2 =10^{-7}\beta_0$, $\eta$ is inversely proportional to
$\gamma_0^2$ when $\gamma_0/\gamma_s < 1$ 
(Fig. \ref{cw-winf}), and it is proportional to $|\epsilon|^2$ [Fig. \ref{epplot}(a)], 
as predicted by the linear theory. 
However, for a relatively large input intensity, i.e., $\kappa_2 > 10^{-4}\beta$, 
the enhancement is reduced.
 This is the result of the nonlinear gain saturation, 
because in our case, gain saturation has a significant effect on the intensity amplification
when $\gamma_0 \ll \sqrt{|\epsilon \kappa_2|}$, 
according to the nonlinear steady-state analysis of
Eqs. (\ref{SB1})--(\ref{SB3}). 
The gain saturation may also change the positions of the EPs in the parametric spaces.
Consequently, the saturation effect limits the enhancement in the
response intensity. 
However, we note that it also makes the enhancement less sensitive 
to the deviation from the EPs (i.e., $|p|>0$) [see Figs. \ref{epplot}(b) and (c)]. 
\subsection{Effects of spontaneous emission noise}
As demonstrated in the previous subsections,  
the reduction in the decay rate $\gamma_0$ by the gain is indispensable 
for enhancing the response intensity.  
However, the presence of the gain inevitably results in spontaneous
emission noise; therefore,
the response to the input waves as well as the spontaneous emission may both be enhanced
if the condition in Eq. (\ref{eq_con}) is satisfied.  
An example of the enhanced spontaneous emission spectrum for 
$\gamma_0 < \gamma_n$ at the EP
is shown in Fig. \ref{fig6}(a). 
The spectrum was calculated by taking into account the noise terms $F_1$
and $F_2$ in Eqs. (\ref{SB1})--(\ref{SB3}) 
with no input waves, i.e., $\kappa_1=\kappa_2=0$. 
The figure also shows the fitting curves, which were obtained 
by comparing the integrals of the numerical spectra and the theoretical 
curves (see Appendix \ref{appen4} for the details).
One can see that the emission spectrum at the EP can be well-fitted by a squared Lorentzian
curve, and the amplitude is greater than that of 
the spectrum measured at the DP. 
The squared Lorentzian shape is general at a second-order EP in any
non-Hermitian photonic system with gain, and it can be experimentally observed. 

Figure \ref{fig6}(b) shows the spectra under the CW wave input
condition at the resonance $\Delta=\Delta_0$. 
One can clearly see that the response intensity to the input wave at the
EP can be more than an order of magnitude greater than that at the DP, 
although the spontaneous emission is also enhanced. 
The signal-to-noise ratio is not degraded in this case.
These results suggest the experimental realization of 
the enhancement in the response to input signals by using EPs. 
  
\section{Summary and discussion \label{secIV}}
The response characteristics of non-Hermitian optical cavities near EPs
were analytically studied on the basis of Eq. (\ref{eq_EPres}). 
Although the $K$-factor (PF) diverges at these points, 
the actual response amplitude is limited to a finite value. 
Importantly, optical cavities operating at an EP can exhibit 
a non-Lorentzian frequency response due to the interference 
between the first and second poles in the resolvent, 
which mainly depends on the decay rate $\gamma_0$, 
the input field $\ff$, and the non-Hermitian system 
parameters $c_{ij}$ [see Eq. (\ref{eq_EPres})].
This is a unique property of systems operating near an EP.  
When a system and the input channel are appropriately designed 
and the decay rate $\gamma_0$ is reduced by gain, 
one can observe a significant enhancement in the response near 
the coalescing resonance at the EP compared to normal systems 
with the same loss and same coupling parameters for the input waves. 
The results of the linear theory were numerically verified in a
microring cavity with non-Hermitian backscattering 
by using a dynamical model taking into account 
the effects of the gain and spontaneous emission.
With the aid of the gain, the intracavity intensity can respond strongly 
to an input field as well as optical noise. 
Although the gain saturation occurring under a strong input field leads to 
a decrease in the response intensity, 
the intensity is still much higher than that of a standard optical
cavity that has the same input wave coupling strength to each eigenmode 
but does not operate at the EP.  
Even when the spontaneous emission is enhanced, 
the signal-to-noise ratio does not decrease because of the excess
response to input signals, as far as a gain saturation effect is not dominant.   
These results suggest the possibility of experimental observation of 
the enhanced response behavior by using EPs.  
For example, an enhancement in a microring cavity operating at an EP 
could be observed in an add-drop configuration with input and output
waveguides
if asymmetric backscattering is so strong that the condition
$\gamma_0 < \gamma_s$ is satisfied.   

A further enhancement is possible if a system can operate at a
higher-order EP because the resolvent or Green's function has
higher-order poles \cite{Lin2016,Heiss2015G}.

The response theory presented in this work is applicable to a variety of
photonic systems, including optical microcavities, 
parity--time symmetric systems,
optomechanical resonators, and plasmonic systems that can operate at an
EP.
The response characteristics at an EP shown in this work, e.g., the
enhancement in or suppression of the response amplitude, 
will be useful for controlling the excitation of eigenmodes, 
the output from resonator sensors at EPs
\cite{WChen2017,Hodaei2017,Ren2017,Sunada2017}, 
extraordinary
optical transmission \cite{HXCui2013}, or the light--matter interactions inside microcavities. 

\begin{acknowledgements}
I would like to thank Professor T. Niyama and Dr. J.-W. Ryu for
 reading the manuscript and offering valuable comments. 
This work was supported by JSPS KAKENHI, Grant No. 16K04974. 
\end{acknowledgements}

\appendix
\section{Derivation of Eq. (\ref{eq_EPres}) \label{appen1}}
The resolvent $\MM_{EP}$ at an EP is derived by assuming 
that $\MM$ can be described by a 2$\times$2 matrix involving
a pair of nearly degenerate modes, $j=1$ and $j=2$. 
Although expressions similar to $\MM_{EP}$ have been reported
 in Ref. \cite{WDHeiss2014,Pick2017,Heiss2015G}, 
the 2$\times$2 matrix form derived in this work allows for 
a simple insight into the relationship between the response
functions and the matrix $\HH$.

Here, let $\Omega_j$ and $\uu_j$ be an eigenvalue and the right eigenvector of mode
$j$, respectively.  
The resolvent $\MM$ is rewritten using an eigenvector matrix
$\PP =  (\uu_1,\uu_2)$ as follows: 
\begin{eqnarray}
\MM(\omega) = \left(\omega\II - \HH \right)^{-1} 
=
\PP\left(\omega\II - \hat{\Lambda} \right)^{-1}\PP^{-1} \nonumber \\ 
=
\left(
\begin{array}{cc}
R_1 + \dfrac{R_1-R_2}{\det\PP}u_{12}u_{21} & 
 -\dfrac{R_1-R_2}{\det\PP}u_{11}u_{12} \\ 
\\
 \dfrac{R_1-R_2}{\det\PP}u_{21}u_{22} &
R_2 - \dfrac{R_1-R_2}{\det\PP}u_{12}u_{21} 
\end{array}
\right), \label{eq_appen1}
\end{eqnarray}
where 
$\hat{\Lambda}$ = diag$\left(\Omega_1,\Omega_2\right)$, 
$R_j = \left(\omega-\Omega_j\right)^{-1}$, and
$\uu_j=(u_{1j},u_{2j})^T$.
Note that $\det\PP$ and $R_1-R_2$ both become zero at an EP, where 
the two eigenvalues coincide, and their 
 eigenvectors completely overlap each other.
The convergence of $(R_1-R_2)/\det\PP$ can be analyzed by 
applying perturbation theory near an EP 
to the above equation \cite{Pick2017}.  
Suppose that $\HH_0$, $\Omega_0$, and $\uu_0 =(u_{10},u_{20})^T$ 
are the Hamiltonian at an EP, the coalesced
eigenvalue, and the corresponding eigenvector, respectively.
According to \cite{Seyranian2003,Seyranian2005}, 
$\Omega_j$ and $\uu_j$ can be expressed as
$\Omega_0\pm\Delta\Omega$ 
and $\uu_j = \uu_0 \pm \Delta\Omega\uu_J + O(|\Delta\Omega|^2)$, 
respectively, near an EP. 
$\Delta\Omega$ is the deviation from $\Omega_0$,
and $\uu_J = (u_{1J},u_{2J})^T$ is 
an associated vector, which satisfies a generalized eigenvalue
equation, $(\HH_0-\Omega_0\II )\uu_J=\uu_0$. 
By using $\uu_0$ and the associated vector $\uu_J$, 
$\det\PP$ is obtained as $-2\Delta\Omega J + O(|\Delta\Omega |^2)$, 
where $J$ is the determinant of the matrix $\PP_J=(\uu_0,\uu_J)$.
We omit the second order of $|\Delta\Omega |^2$ near an EP 
and obtain 
\begin{eqnarray}
\MM(\omega)
\approx 
\left(
\begin{array}{cc}
R_a + c_{11} R_d & 
 c_{12} R_d \\
\\
 c_{21} R_d &
R_a +c_{22} R_d
\end{array}
\right),
\end{eqnarray}
where 
$R_a=(R_1+R_2)/2$ and
\begin{eqnarray}
R_d &=&  \dfrac{1}{2\Delta\Omega}
\left(
\dfrac{1}
{\omega-\Omega_1} 
- \dfrac{1}{\omega - \Omega_2}
\right) \nonumber \\ 
&=&
\dfrac{1}
{\left(
\omega - \Omega_0
\right)^2
-\Delta\Omega^2
}   
. \label{eq_Rd}  
\end{eqnarray}
When $\Delta\Omega \rightarrow 0$, $R_a\rightarrow R_0$, and 
$R_d \rightarrow R_0^2$. 
In the above, $c_{11}=-u_{10}u_{20}/J$, 
$c_{12}= u_{10}^2/J$, 
$c_{21}=-u_{20}^2/J$, and $c_{22}=-c_{11}$. 
These coefficients correspond to the matrix components of 
$(\HH_0-\Omega_0\II)$.
This can be easily confirmed by calculating $(\HH_0-\Omega_0\II)
=\PP_J(\hat{T}_0-\Omega_0\II)\PP_J^{-1}$, 
where 
\begin{eqnarray}
\hat{T}_0 = 
\left(
\begin{array}{cc}
\Omega_0& 
 1\\
\\
 0 &
\Omega_0
\end{array}
\right).
\end{eqnarray}

\begin{figure}
\begin{center}   
  \begin{tabular}{c}
\hspace*{-0.4cm}
\raisebox{0.0cm}{\includegraphics[width=6cm]{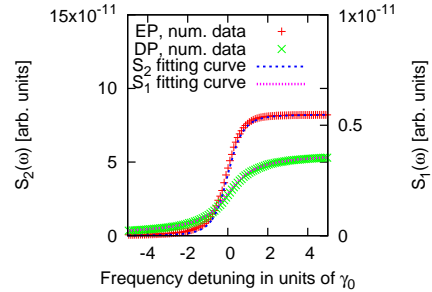}}
  \end{tabular}
\end{center}
\vspace{-4mm}
\caption{\label{fig7} 
Integrals of the noise spectra shown in Fig. \ref{fig6}(a). 
The integrals of the spectra at the EP and DP are 
fitted by $S_{1}$ and $S_2$, respectively.}
\end{figure}

\section{Fitting of the spontaneous emission spectra \label{appen4}}
Because the emission spectra shown in Fig. \ref{fig6}(a) fluctuate,
the integrals of the spectra are compared to those of 
the theoretical spectra.  
This fitting method is similar to that used in
Ref. \cite{Cerjan2015}.
According to the linear theory presented in Sec. \ref{secII}, 
the emission spectrum at the EP is represented by 
a squared Lorentzian curve when $\gamma_0 \ll \gamma_n$, 
whereas the spectrum at the DP ($\epsilon=p=0$) is a Lorentzian curve. 
The integral of a Lorentzian curve, $L(\omega) = A_1/[\gamma_0^2 +
(\omega-\Delta_0)^2]$, is 
 \begin{eqnarray}
S_1(\omega) = \int_{-\infty}^{\omega} L(\omega)d\omega 
=
\dfrac{A_1}{\gamma_0}\left[
\theta(\omega)
+ 
\dfrac{\pi}{2}
\right],
\end{eqnarray}
where $A_1$ is a fitting parameter, and $\theta(\omega) =
\tan^{-1}[(\omega-\Delta_0)/\gamma_0]$. 
The integral of a squared Lorentzian curve, $L_2(\omega) = A_2|\epsilon|^2/(4[\gamma_0^2 +
(\omega-\Delta_0)^2]^2)$, is 
 \begin{eqnarray} 
S_2(\omega) &=& \int_{-\infty}^{\omega} L_2(\omega)d\omega \nonumber \\
&=&
\dfrac{A_2|\epsilon|^2}{8\gamma_0^3}\left[
\theta(\omega)
+
\dfrac{1}{2}\sin
2\theta(\omega)
+ 
\dfrac{\pi}{2}
\right],
\end{eqnarray}
where $A_2$ is a fitting parameter.
The fitting results are shown in Fig. \ref{fig7}. 


\begin{thebibliography}{99}
\bibitem{Okolowicz2003}
J. Okolowicz, M. Ploszajczak, and I. Rotter,
Phys. Rep. {\bf 374}, 271 (2003).
%

\bibitem{Trefethen2005}
L. N. Trefethen and M. Embree,
{\it Spectra and Pseudospectra: The Behavior of Nonnormal Matrices and
	Operators}
(Princeton University Press, Princeton, NJ, 2005).

\bibitem{Trefethen1993}
L. N. Trefethen, A. E. Trefethen, S. C. Reddy and T. A. Driscoll,
Science {\bf 261}, 578-584 (1993). 

\bibitem{Farrell1994}
Brian F. Farrell and Petros J. Ioannou,
Phys. Rev. Lett. {\bf 72} 1188 (1994).
 
\bibitem{Szirmai2009}
G. Szirmai, D. Nagy, and P. Domokos,
Phys. Rev. Lett. {\bf 102}, 080401 (2009). 

\bibitem{Papoff2008}
F. Papoff, G. D'Alessandro, and G.-L. Oppo,
Phys. Rev. Lett. {\bf 100}, 123905 (2008).

\bibitem{Longhi2000}
S. Longhi and P. Laporta,
Phys. Rev. E {\bf 61}, R989(R) (2000).

\bibitem{Alessandro2009}
G. D'Alessandro and F. Papoff,
Phys. Rev. A {\bf 80}, 023804 (2009). 

\bibitem{Biancalani2017}
Tommaso Biancalani, Farshid Jafarpour, and Nigel Goldenfeld,
Phys. Rev. Lett. {\bf 118}, 018101 (2017).

\bibitem{Petermann1979}
K. Petermann, 
IEEE J. Quantum Electron. {\bf QE-15}, 566 (1979).

\bibitem{Haus1985}
H. A. Haus and S. Kawakami,
IEEE J. Qunatum Electron. {\bf 21}, 63 (1985). 

\bibitem{Siegman1989-1}
A. E. Siegman,
Phys. Rev. A {\bf 39}, 1253 (1989); 
%
{\bf 39}, 1264 (1989). 

\bibitem{Siegman1995}
A. E. Siegman, 
Appl. Phys. B {\bf 60}, 247 (1995). 

\bibitem{Exter2001}
M. P. van Exter, N. J. van Druten, A. M. van der Lee, S. M. Dutra,
	G. Nienhuis, and J. P. Woerdman,
Phys. Rev. A {\bf 63}, 043801 (2001). 

\bibitem{Cheng2006}
Yuh-Jen Cheng,
Phys. Rev. Lett. {\bf 97}, 093601 (2006). 

\bibitem{SYLee2008}
S.-Y. Lee, J.-W. Ryu, J.-B. Shim, S.-B. Lee, S. W. Kim, and K. An,
Phys. Rev. A {\bf 78}, 015805 (2008). 

\bibitem{Cheng1996}
Yuh-Jen Cheng, C. G. Fanning, and A. E. Siegman,
Phys. Rev. Lett. {\bf 77}, 627 (1996). 

\bibitem{Eijkelenborg1996}
M. A. van Eijkelenborg, A. M. Lindberg, M. S. Thijssen, and
	J. P. Woerdman,
Phys. Rev. Lett. {\bf 77}, 4314 (1996).

\bibitem{Yao1992}
Gang Yao, Y. C. Chen, C. M. Harding, S. M. Sherrick, R. J. Dalby,
	R. G. Waters, and C. Largent,
Opt. Lett. {\bf 17}, 1207-1209 (1992). 

\bibitem{vanderLee1997}
A. M. van der Lee, N. J. van Druten, A. L. Mieremet, M. A. van
	Eijkelenborg, A. M. Lindberg, M. P. van Exter, and
	J. P. Woerdman,
Phys. Rev. Lett. {\bf 79}, 4357 (1997). 

\bibitem{vanderLee2000}
A. M. van der Lee, A. L. Mieremet, M. P. van Exter, N. J. van Druten,
	and J. P. Woerdman,
Phys. Rev. A {\bf 61}, 033812 (2000). 

\bibitem{Kato1966}
T. Kato, {\it Perturbation theory for Linear Operators} 
(Springer, New York, 1966). 

\bibitem{Berry2004}
M. V. Berry, 
Czech. J. Phys. {\bf 54}, 1039-1047 (2004).

\bibitem{Heiss2012}
W. D. Heiss, 
J. Phys. A: Math. Theor. {\bf 45}, 444016 (2012).

\bibitem{Cao2015}
H. Cao and J. Wiersig, 
Rev. Mod. Phys. {\bf 87}, 61-111 (2015).

\bibitem{SBLee2009}
S.-B. Lee, J. Yang, S. Moon, S.-Y. Lee, J.-B. Shim, S. W. Kim, 
J.-H. Lee, and K. An,
Phys. Rev. Lett. {\bf 103}, 134101 (2009). 

\bibitem{Zhen2015}
B. Zhen, C. W. Hsu, Y. Igarashi, L. Lu, I. Kaminer,
A. Pick, S.-L. Chua, J. D. Joannopoulos, and M. Solja$\check{c}$i$\acute{c}$,
Nature {\bf 525}, 354-358 (2015).
%
\bibitem{Kim2016}
K.-H. Kim, M.-S. Hwang, H.-R. Kim, J.-H. Choi, Y.-S. No, 
and H.-G. Park,
Nat. Commun. {\bf 7} 13893 (2016).

\bibitem{Doppler2016}
J. Doppler, A. A. Mailybaev, J. B{\"o}hm, U. Kuhl, A. Girschik, F. Libisch, T. J. Milburn,
P. Rabl, N. Moiseyev and S. Rotter,
Nature {\bf 537}, 76-79 (2016).

\bibitem{Liertzer2012}
M. Liertzer, Li Ge, A. Cerjan, A. D. Stone, H. E. T{\"u}reci, and
	S. Rotter,
Phys. Rev. Lett. {\bf 108}, 173901 (2012).

\bibitem{BoPeng2015}
B. Peng, $\c{S}$. K. {\"O}zdemir, S. Rotter, H. Yilmaz, 
M. Liertzer, F. Monifi, C. M. Bender, F. Nori, and L. Yang,
Science {\bf 346}, 328 (2015). 

\bibitem{BoPeng2014}
B. Peng, $\c{S}$. K. {\"O}zdemir, F. Lei, F. Monifi, M. Gianfreda,
G. L. Long, S. Fan, F. Nori, C. M. Bender, and L. Yang,
Nat. Phys. {\bf 10}, 394 (2014). 

\bibitem{LChang2014}
L. Chang, X. Jiang1, S. Hua1, C. Yang, J. Wen, L. Jiang,
G. Li, G. Wang, and M. Xiao,
Nat. Photon. {\bf 8}, 524 (2014). 

\bibitem{Lin2013}
Z. Lin, H. Ramezani, T. Eichelkraut, T. Kottos, H. Cao, and
	D. N. Christodoulides,
Phys. Rev. Lett. {\bf 106}, 213901 (2011). 

\bibitem{Feng2013}
L. Feng, Y.-L. Xu, W. S. Fegadolli, M.-H. Lu, J. E. B. Oliveira, 
V. R. Almeida, Y.-F. Chen, and A. Scherer, 
Nat. Mater. {\bf 12}(2), 108 (2013). 

\bibitem{Wiersig2014}
J. Wiersig,
Phys. Rev. Lett {\bf 112}, 203901 (2014);
J. Wiersig, 
Phys. Rev. A {\bf 93}, 033809 (2016).

\bibitem{WChen2017}
W. Chen, $\c{S}$. K. {\"O}zdemir, G. Zhao, J. Wiersig, and L. Yang,
Nature {\bf 548}, 192 (2017).

\bibitem{Hodaei2017}
H. Hodaei, A. U. Hassan, S. Wittek, H. Garcia-Gracia, R. El-Ganainy,
D. N. Christodoulides, and M. Khajavikhan,
 Nature, {\bf 548}(7666) 187 (2017). 

\bibitem{Ren2017}
J. Ren, H. Hodaei, G. Harari, A. U. Hassan, W. Chow, M. Soltani,
D. N. Christodoulides, and M. Khajavikhan, 
Opt. Lett. {\bf 42}, 1556 (2017).

\bibitem{Sunada2017}
S. Sunada,
Phys. Rev. A {\bf 96}, 033842 (2017). 

\bibitem{Demmel1997}
J. Demmel, {\it Applied Numerical Linear Algebra}
(SIAM, Philadelphia, 1997).
%
\bibitem{Comment11}
In other words, $K_j$ is 1 only for normal systems, where 
the set of the eigenvectors forms an orthogonal basis, and 
$\uu_j$ can be parallel to $\vv_j$, e.g., see \cite{Demmel1997}.

\bibitem{Seyranian2003}
A. P. Seyranian and A. A. Mailybaev,
{\it Multiparameter Stability Theory with Mechanical Applications} 
(Singapore: World Scientific, 2003). 

\bibitem{Seyranian2005}
A. P. Seyranian, O. N. Kirillov and A. A. Mailybaev,
J. Phys. A: Math. Gen. {\bf 38}, 1723 (2005). 

\bibitem{Pick2017}
A. Pick, B. Zhen, O. D. Miller, C. W. Hsu, F. Hernandez,
A. W. Rodriguez, M. Solja$\check{c}$i$\acute{c}$, and S. G. Johnson, 
Opt. Express {\bf 25}(11), pp. 12325-12348 (2017). 

\bibitem{WDHeiss2014}
W. D. Heiss and G. Wunner,
Eur. Phys. J. D {\bf 68}, 284 (2014). 

\bibitem{Suh2004}
W. Suh, Z. Wang, and S. Fan,
IEEE J. Quantum Electron. {\bf 40}, 1511 (2004). 

\bibitem{Lin2016}
Z. Lin, A. Pick, M. Lon$\breve{c}$ar, and A. W. Rodriguez, 
Phys. Rev. Lett. {\bf 117}, 107402 (2016).

\bibitem{Yoo2011}
G. Yoo, H.-S. Sim, and H. Schomerus,
Phys. Rev. A {\bf 84}, 063833 (2011). 

\bibitem{Peng2016}
B. Peng, S. K. {\"O}zdemir, M. Liertzer, W. Chen, J. Kramer, H. Yilmaz,
J. Wiersig, S. Rotter, and L. Yang ,
PNAS {\bf 113}, 6845 (2016).

\bibitem{Wiersig2011}
J. Wiersig, A. Eberspacher, J.-B. Shim, J.-W. Ryu, S. Shinohara,
	M. Hentschel, and H. Schomerus
Phys. Rev. A {\bf 84}, 023845 (2011). 

\bibitem{Shu2016}
F.-J. Shu, C.-L. Zou, X.-B. Zou, and L. Yang,
Phys. Rev. A {\bf 94}, 013848 (2016). 

\bibitem{Farrell1996}
B. F. Farrell and P. J. Ioannou,
J. Atmos. Sci. {\bf 53}, 2025 (1996). 

\bibitem{Harayama2005}
T. Harayama, S. Sunada, and K. S. Ikeda,
Phys. Rev. A {\bf 72}, 013803 (2005).

\bibitem{Sunada2013}
S. Sunada, T. Fukushima, S. Shinohara, T. Harayama, and M. Adachi,
Phys. Rev. A {\bf 88}, 013802 (2013).

\bibitem{Drummond1991}
P. D. Drummond and M. G. Raymer,
Phys. Rev. A {\bf 44}, 2072 (1991). 

\bibitem{Cerjan2015}
A. Cerjan, A. Pick, Y. D. Chong, S. G. Johnson, and A. D. Stone,
Opt. Express {\bf 23}, 28316 (2015). 

\bibitem{Heiss2015G}
W. D. Heiss, 
Int. J. Theor. Phys. {\bf 54} 3954 (2015). 

\bibitem{HXCui2013}
H-X. Cui, X.-W. Cao, M. Kang, T.-F. Li, M. Yang, T.-J. Guo, Q.-H. Guo,
	and J. Chen,
Opt. Express {\bf 21}, 13368 (2013). 

\end{thebibliography}
\end{document}